\documentstyle[preprint,prl,aps]{revtex} 

\newcommand{\Tr}{{\rm Tr}\,}
\newcommand{\CD}{{\cal D}}
\newcommand{\CN}{{\cal N}}

\newcommand{\CM}{{\cal M}}
\newcommand{\CP}{{\cal P}}
\newcommand{\CF}{{\cal F}}
\newcommand{\CZ}{{\cal Z}}

\newcommand{\CS}{{\cal S}}

\newcommand{\Pf}{{\rm Pf}}

\newcommand{\half}{\frac{1}{2}}
\font \barfont = bbm10

\newcommand{\id}{\mbox{\bf\barfont 1}} 
\begin{document}
\draft
\title{\begin{flushright}
{\small\hfill AEI-2000-022\\
\hfill hep-th/0004076}\\
\end{flushright}
Yang-Mills Integrals for Orthogonal, Symplectic and
Exceptional Groups}
\author{Werner Krauth \footnote{krauth@physique.ens.fr }}
\address{CNRS-Laboratoire de Physique Statistique,
Ecole Normale Sup\'{e}rieure\\
24, rue Lhomond\\ F-75231 Paris Cedex 05, France}
\author{Matthias Staudacher \footnote{matthias@aei-potsdam.mpg.de }
\footnote{Supported in part by EU Contract FMRX-CT96-0012} }
\address{
Albert-Einstein-Institut, Max-Planck-Institut f\"{u}r
Gravitationsphysik\\ Am M\"uhlenberg 1\\  D-14476 Golm, Germany }
\maketitle
\begin{abstract}

We apply numerical and analytic techniques to the study
of Yang-Mills integrals with orthogonal, symplectic and exceptional
gauge symmetries. The main focus is on the supersymmetric
integrals, which correspond essentially to the bulk part of the
Witten index for susy quantum mechanical gauge theory.  We evaluate
these integrals for $D=4$ and group rank up to three,
using Monte Carlo methods. Our results are at variance with previous
findings. We further compute the integrals with
the deformation technique of Moore, Nekrasov and Shatashvili, which
we adapt to the groups under study. Excellent agreement with all our
numerical calculations is obtained. We also discuss the
convergence properties of the purely bosonic integrals.

\end{abstract}
\vspace{5.5cm}
\newpage
\narrowtext

\section{Introduction}

Recent attempts to describe $D$-branes through effective actions
have revealed the existence of a new class of gauge-invariant matrix
models. These models are related to ($-1$)-branes. They differ
from the classic systems of random matrices, which have been
extensively studied ever since Wigner's and Dyson's work in the
1950's.  The models consist in matrix integrals of $D$ non-linearly
coupled matrices.  They are obtained by complete dimensional
reduction of $D$-dimensional Euclidean continuum (susy) Yang-Mills
theory to zero dimensions, and we term them quite generally {\it
Yang-Mills integrals}.

The supersymmetric $D=4,6,10$ integrals with SU$(N)$ symmetry have
already found several important applications.  They are relevant
to the calculation of the Witten index of quantum mechanical gauge
theory \cite{smilga},\cite{sestern}, and to multi-instanton
calculations \cite{hollo1},\cite{hollo2} of four-dimensional
SU$(\infty)$ susy conformal gauge theory.  The $D=10$ integrals
are furthermore the crucial ingredient in the so-called IKKT model
\cite{ikkt1}, which possibly provides a non-perturbative definition
of IIB superstring theory.  Finally, it remains to be seen whether
Yang-Mills integrals contain information on the full, unreduced
field theory through the Eguchi-Kawai mechanism \cite{ek} as the
size of the matrices gets large. Some very interesting recent
considerations along these lines can be found in \cite{aabhn}.

Yang-Mills integrals are ordinary, not functional integrals.
Despite this tremendous simplification, no systematic analytic
tools for their investigation are known to date.  We have developed
\cite{kns},\cite{ks1},\cite{ks2},\cite{kps} accurate and reliable
Monte Carlo methods which allow to study the new matrix models as
long as the dimension of  the gauge group is not too large.  We
have found, e.g., that supersymmetry is generically {\it not}
necessary for the existence of the integrals \cite{ks1}.  We also
computed their asymptotic eigenvalue distributions, which we found
to qualitatively differ between the susy and bosonic case as the
size of the gauge group gets large \cite{ks2}.  For related,
complementary studies see \cite{nishi}.

To date all existing studies have focused on the case where the
gauge group is SU$(N)$. In the present paper we generalize to the
cases of all other semi-simple compact Lie groups of rank $r\leq
3$.  These are, (besides the  already known cases SU$(2)$, SU$(3)$,
SU$(4)$) the groups SO$(3)$, SO$(4)$, SO$(5)$, SO$(6)$, SO$(7)$,
Sp$(2)$, Sp$(4)$, Sp$(6)$, and G$_2$, for which we compute the susy
$D=4$ partition functions.  We were motivated in part by a recent paper
of Kac and Smilga \cite{kasm} which presented conjectures about
the values of the bulk part of the Witten index (and therefore for
the corresponding integrals).  Intriguingly, our results are at
variance with their predictions in most cases, indicating that the
index calculations for these groups are even more subtle than the
corresponding considerations for SU$(N)$, where the approach of
\cite{kasm} agrees with the known values.
 
Moore, Nekrasov and Shatashvili \cite{mns}  recently employed sophisticated
deformation techniques to evaluate 
the SU$(N)$ susy bulk index for all $N$ and $D=4,6,10$.  The method
apparently leads to the correct result for all SU$(N)$ \cite{kns},
\cite{ks1}.  Below, we  adapt the technique to the more general groups, 
and we find again excellent agreement with the  Monte Carlo
calculation. This further indicates that the deformation method is
indeed reliable.

\section{Yang-Mills integrals for semi-simple compact gauge groups}

For a general semi-simple compact Lie group $G$ we define
supersymmetric or bosonic Yang-Mills integrals as
\begin{equation}
\CZ^{\CN}_{D,G}:=\int \prod_{A=1}^{{\rm dim}(G)} 
\Bigg( \prod_{\mu=1}^{D} \frac{d X_{\mu}^{A}}{\sqrt{2 \pi}} \Bigg) 
\Bigg( \prod_{\alpha=1}^{\CN} d\Psi_{\alpha}^{A} \Bigg)
\exp \bigg[  \frac{1}{4 g^2} \Tr 
[X_\mu,X_\nu] [X_\mu,X_\nu] + \frac{1}{2 g^2} 
\Tr \Psi_{\alpha} [ \Gamma_{\alpha \beta}^{\mu} X_{\mu},\Psi_{\beta}]
\bigg],
\label{susyint}
\end{equation}
where dim$(G)$ is the dimension of the Lie group and the $D$ bosonic
matrices $X_{\mu}=X_{\mu}^A T^A$ and the $\CN$ fermionic matrices
$\Psi_{\alpha}=\Psi_{\alpha}^A T_A$ 
are anti-hermitean and take values in the
fundamental representation of the Lie algebra Lie$(G)$,
whose generators we denote by $T^A$. The integral eq.(\ref{susyint})
depends on the gauge coupling constant in a trivial fashion,
as we can immediately scale out $g$. Nevertheless, there is a natural
convention for fixing $g$: For an orthogonal set of
generators we should pick $g$ according to their normalization:
$\Tr T^A T^B=-g^2 \delta^{AB}$. This convention is imposed by
the index calculations of the next section.
In the bosonic case we simply drop the fermionic variables. 

The supersymmetric integrals can formally be defined in $D=3,4,6,10$, 
which corresponds to $\CN=2,4,8,16$ real supersymmetries. 
We are not aware of a mathematically rigorous investigation of their
convergence properties. However,
our numerical studies indicate that they are absolutely convergent
in $D=4,6,10$ (but not in $D=3$) for all semi-simple compact
gauge groups. The convergence properties of the bosonic ($\CN=0$)
integrals are discussed in section 6. The variables
$\Psi_{\alpha}^{A}$ are real Grassmann-valued and can be integrated out,
leading to a bosonic integral with very special measure:
\begin{equation}
\CZ^{\CN}_{D,G} = \int \prod_{A=1}^{{\rm dim}(G)} \prod_{\mu=1}^{D}  
\frac{d X_{\mu}^{A}}{\sqrt{2 \pi}}
\exp \bigg[  \frac{1}{4 g^2} \Tr
[X_\mu,X_\nu] [X_\mu,X_\nu] \bigg]~
\CP_{D,G}(X).
\label{int}
\end{equation}
$\CP_{D,G}(X)$ is a homogeneous Pfaffian polynomial of degree 
$\half \CN~{\rm dim}(G)$ given by
\begin{equation}
\CP_{D,G}=\Pf \CM_{D,G}
\qquad {\rm with} \qquad 
\big(\CM_{D,G}\big)_{\alpha \beta}^{A B}=-i f^{ABC} \Gamma^\mu_{\alpha \beta}
X_\mu^C,
\label{pfaff}
\end{equation}
where the structure constants
\footnote{Note that in\cite{kns} we used hermitean generators
but defined the structure constants through 
$\Big[ T^A,T^B \Big]=i f^{ABC} T^C$, so eq.(\ref{pfaff}) remains
valid.}  
are defined through the real Lie algebra
$\Big[ T^A,T^B \Big]=f^{ABC} T^C$. Explicit expressions for
the Gamma matrices $\Gamma^\mu_{\alpha \beta}$ and further
details on $\CP_{D,G}(X)$ may be found in \cite{kns}.  

\section{Group volumes and bulk indices} 

It is well known that the susy Yang-Mills integrals eq.(\ref{int})
naturally appear when one computes the Witten index of quantum
mechanical gauge theory (i.e.~the reduction of the field theory
to one, as opposed to zero, dimension) by the heat kernel method.
For the details of the method we refer to \cite{smilga},\cite{sestern}. 
Specifically, the integrals are related to the 
bulk part ind$^D_0(G)$ of the index as
\begin{equation}
{\rm ind}_0^D(G)={\rm lim}_{\beta \rightarrow 0} \Tr (-1)^F e^{-\beta H}=
\frac{1}{\CF_G} \CZ^{\CN}_{D,G}.
\label{bind}
\end{equation}
The total Witten index ``ind$^D(G)$'' is then the sum of this bulk part and
a boundary contribution ``ind$^D_1(G)$'': 
${\rm ind}^D(G)={\rm ind}^D_0(G)+{\rm ind}^D_1(G)$.
The constant $\CF_G$ relating the bulk index ind$^D_0$ and the
Yang-Mills integral is independent of $D$ and can be interpreted as
\begin{equation}
\CF_G=\frac{1}{(2 \pi)^{\half {\rm dim}(G)}}~
{\rm Volume}\Big[\frac{G}{Z_G}\Big],
\label{volume}
\end{equation}
i.e.~essentially the volume of the true gauge group, which turns
out to be the quotient group $G/Z_G$, with $Z_G$ the center group of $G$.
In practice, great care has to be taken in using the relation
eq.(\ref{volume}), as the volume depends on the choice of the local
metric on the group manifold.  For the present purposes we simply
adapted our method for computing $\CF_{{\rm SU}(N)}$ (see \cite{ks1})
to the relevant gauge groups.  An invariant average over the group
allows to project onto gauge invariant states and to derive
eq.(\ref{bind}) from the quantum mechanical path integral.  In the
ultralocal limit, the quantum mechanics of $D-1$ matrices turns
into an integral over $D$ matrices. Then, this integration is over
the anti-hermitean generators of the group
%
%
\begin{equation}
\CD U \rightarrow \frac{1}{\CF_G} \prod_{A=1}^{{\rm dim} (G)} 
\frac{d X^A_D}{\sqrt{2 \pi}}.
\label{haarflat}
\end{equation}
The normalized Haar measure $\CD U$ on the group elements 
$U \in G$ simplifies
significantly if we restrict attention to the Cartan subgroup
of $G$. A beautiful result of Weyl allows to explicitly 
write down the restricted measure. If we parametrize the Cartan
torus $T$ by angles $-\pi\leq \theta_1 \leq \pi$, $\ldots$, 
$-\pi\leq \theta_r \leq \pi$, where $r=$rank$(G)$, the measure
reads
\begin{equation}
\CD U \rightarrow \CD T=\frac{1}{|W_G|}
~\Bigg( \prod_{i=1}^r \frac{d \theta_i}{2 \pi} \Bigg)
~|\Delta_G|^2,
\label{weyl}
\end{equation}
where $|W_G|$ is the order of the Weyl group $W_G$ of $G$, and
$|\Delta_G|^2$ the squared modulus of the Weyl denominator:
\begin{equation}
\Delta_G=\prod_{\alpha > 0} \Big[
e^{\frac{i}{2} (\theta,\alpha)}-e^{-\frac{i}{2} (\theta,\alpha)} \Big].
\end{equation}
Here the product is over the set of positive roots of the Lie algebra
Lie$(G)$. In the vicinity of the identity in $G$ the angles $\theta_i$
are small and we can approximate the measure eq.(\ref{weyl}) by
\begin{equation}
\frac{1}{|W_G|}~\Bigg( \prod_{i=1}^r \frac{d \theta_i}{2 \pi} \Bigg)
~\prod_{\alpha>0} \Big[
\half (\theta,\alpha)-\half (\theta,\alpha) \Big]^2.
\label{approx}
\end{equation}
Now restricting the flat measure on Lie$(G)$ on the right hand side of 
eq.(\ref{haarflat}) to the Cartan modes $\theta_i$ we get
\begin{equation}
\prod_{A=1}^{{\rm dim} (G)} 
\frac{d X^A_D}{\sqrt{2 \pi}}
\rightarrow \frac{\CF_G}{(2 \pi)^r} \frac{|Z_G|}{|W_G|}
~\Big( \prod_{i=1}^r d \theta_i \Big)
~\prod_{\alpha>0} \Big[
\half (\theta,\alpha)-\half (\theta,\alpha) \Big]^2.
\label{reduction}
\end{equation}
An important subtlety is that we needed to multiply the measure
eq.(\ref{reduction}) by an additional factor $|Z_G|$ of the order 
of the center group $Z_G$ of $G$, as the averaging over the group manifold
localizes on $|Z_G|$ points. 
Finally, the constant $\CF_G$ in eq.(\ref{reduction}) is fixed 
by noting that the flat measure on Lie$(G)$ is normalized with
respect to Gaussian integration
\begin{equation}
\int \Bigg( \prod_{A=1}^{{\rm dim} (G)}  
\frac{d X^A_D}{\sqrt{2 \pi}}\Bigg)~\exp\Big[-\half \sum_A (X_D^A)^2\Big]=1.
\label{gauss}
\end{equation}
The Gaussian integration of the right hand side of eq.(\ref{reduction})
leads to Selberg-type integrals, see e.g.~\cite{mehta}.
Explicit details on how to implement the above procedure for
the groups under study can be found in the appendices.
With our conventions for the normalization of the generators 
(SO$(N)$: $\Tr T^A T^B=-2 \delta^{AB}$ and Sp$(2N)$, G$_2$: 
$\Tr T^A T^B=-\half \delta^{AB}$) one finds
\begin{equation}
\CF_{{\rm SO}(N)}=\frac{1}{2 C_N} 
~{\pi^{\frac{N}{2}} \over 2^{\frac{N(N-5)}{4}}~\prod_{j=1}^N \Gamma(j/2)},
\label{groupfacso}
\end{equation}
where $C_{2 N}=2$ and $C_{2 N+1}=1$, as well as
\begin{equation}
\CF_{{\rm Sp}(2 N)}=\frac{1}{2} 
~{2^{2 N^2+\frac{N}{2}}~\pi^{\frac{N}{2}} \over \prod_{j=1}^N \Gamma(2 j)}
\label{groupfacsp}
\end{equation}
and finally
\begin{equation}
\CF_{{\rm G}_2}=\frac{36864~\sqrt{3}~\pi}{5}.
\label{groupfacg}
\end{equation}

\section{Deformation method}

In \cite{mns} it was suggested that the original susy Yang-Mills
integrals eq.(\ref{susyint}) may be vastly simplified by a deformation
technique. It consists in adding a number of terms to the action which
break the number of supersymmetries from $\CN=2,4,8,16$ to
$\CN=1$. Keeping one of the supersymmetries means that the partition
function is ``protected'' and 
should not change under the deformation. This gives the
correct result\footnote{For unclear reasons it fails
for $D=3$.} for SU$(N)$ and $D=4,6,10$. The final outcome is
a much simpler integral involving only a single Lie-algebra valued
matrix. The remaining integral is still invariant under the gauge group,
and one can therefore pass from the full algebra to the Cartan
subalgebra degrees of freedom. This was derived in \cite{mns} in detail
for SU$(N)$ but should carry over immediately to other gauge groups. 
For $D=4$ ($\CN=4$) one finds, in the notation of the previous
section (here the product is over all roots $\alpha$ of Lie$(G)$)
\begin{equation}
{\rm ind}_0^{D=4}(G)=
\frac{|Z_G|}{|W_G|}~\frac{1}{E^r}
\int~\Bigg( \prod_{i=1}^r \frac{d x_i}{2 \pi i} \Bigg)
~\prod_{\alpha} \frac{\Big[\half (x,\alpha)-\half (x,\alpha) \Big]}
{\Big[\half (x,\alpha)-\half (x,\alpha) -E \Big]}.
\label{contour}
\end{equation}
This $r$-dimensional integral ($r=$rank$(G)$) is divergent. 
There are divergences due to the poles of the denominator of
eq.(\ref{contour}), as well as at infinity, where the integrand
tends to one. These divergences are present since the method starts
from the partition sums eq.(\ref{susyint}) with Minkowski signature, 
i.e.~with the Wick-rotated versions of our integrals. The Minkowski integrals 
are divergent without a prescription. The poles are regulated by
giving an imaginary part to the parameter $E$. The singularity at
infinity is regulated by interpreting the integrals as {\it contour}
integrals\footnote{This interpretation furthermore necessitates the
inclusion of the factors of $i$ in the measure of eq.(\ref{contour})
which would not be present in an ordinary integral over a real
Lie algebra.}. It would be interesting to complete the arguments by
demonstrating that the Wick-rotation leads to precisely these
prescriptions. Very encouraging signs for the consistency of
this method are that the final result neither depends on the location
of the parameter $E$ nor on whether the contours are closed
in the upper or the lower half plane (it is important though that
all $r$ contours are closed in the same way). For $D=6,10$
expressions very
similar to eq.(\ref{contour}) can be found in \cite{mns}.

We now present the explicit form of the contour integrals eq.(\ref{contour})
for the groups studied in the present work (see appendices for
details) 

\[
{\rm ind}_0^{D=4}({\rm SO}(2N+1))=
\frac{1}{2^N N!}~\frac{1}{E^N}
~\oint \prod_{i=1}^N \frac{d x_i}{2 \pi i}
~\prod_{i<j}^N
\frac{(x_i^2-x_j^2)^2}{\Big[(x_i-x_j)^2-E^2\Big] \Big[(x_i+x_j)^2-E^2\Big] }
\times
\]
\begin{equation}
\makebox[3.3cm]{} \times \prod_{i=1}^N \frac{x_i^2}{x_i^2-E^2}
\label{soodd}
\end{equation}
\begin{equation}
{\rm ind}_0^{D=4}({\rm SO}(2N))=
\frac{2}{2^{N-1} N!}~\frac{1}{E^N}
~\oint \prod_{i=1}^N \frac{d x_i}{2 \pi i}
~\prod_{i<j}^N
\frac{(x_i^2-x_j^2)^2}{\Big[(x_i-x_j)^2-E^2\Big] \Big[(x_i+x_j)^2-E^2\Big] }
\label{soeven}
\end{equation}
\[
{\rm ind}_0^{D=4}({\rm Sp}(2N))=
\frac{2}{2^N N!}~\frac{1}{E^N}
~\oint \prod_{i=1}^N \frac{d x_i}{2 \pi i}
~\prod_{i<j}^N
\frac{(x_i^2-x_j^2)^2}{\Big[(x_i-x_j)^2-E^2\Big] \Big[(x_i+x_j)^2-E^2\Big] }
\times
\]
\begin{equation}
\makebox[2.8cm]{} \times \prod_{i=1}^N \frac{x_i^2}{x_i^2-(\frac{E}{2})^2}
\label{sp}
\end{equation}
\vspace{3cm}
\[
{\rm ind}_0^{D=4}({\rm G}_2)=
\frac{1}{12}~\frac{1}{E^2}
~\oint \frac{d x_1}{2 \pi i} \frac{d x_2}{2 \pi i}
\frac{(x_1-x_2)^2 (x_1+x_2)^2 x_1^2 x_2^2 } 
{\Big[(x_1-x_2)^2-E^2 \Big]
\Big[ (x_1+x_2)^2-E^2 \Big]  \Big[ x_1^2-E^2 \Big] \Big[ x_2^2-E^2 \Big]} 
\times 
\makebox[5cm]{}
\] 
\begin{equation}
\makebox[2.8cm]{} \times \frac{ (2x_1+x_2)^2 (x_1+2x_2)^2} 
{ \Big[(2 x_1 +x_2)^2-E^2 \Big] \Big[ (x_1+2x_2)^2-E^2 \Big]}
\label{g}
\end{equation}

They are easily evaluated for low rank, and we present the results in
table 1.  We highlighted the cases which were not already indirectly known
due to the standard low-rank isomorphisms 
${\rm so}(3)={\rm sp}(2)={\rm su}(2)$,
${\rm so(4)}={\rm su}(2) \oplus {\rm su}(2)$, ${\rm so}(6)={\rm su}(4)$.
Note, however, that these identities, as well as the final semi-simple 
Lie algebra isomorphism ${\rm so}(5)={\rm sp}(4)$, constitute non-trivial
consistency checks on the expression eq.(\ref{contour}), as the 
precise form of the
corresponding contour integrals is different in all these cases. 
It would be interesting to compute eqs.
(\ref{soodd}),(\ref{soeven}),(\ref{sp}),(\ref{g}) for arbitrary rank,
as has been done in the case of SU$(N)$ in \cite{mns}. 
We also checked that the analogous, more complicated $D=6$ contour
integrals lead to the same bulk indices, as one expects.

For the groups not related by an isomorphism to SU$(N)$ the 
rational numbers in table~1 differ from the ones
proposed in \cite{kasm}. 
I would be important to understand why.
We also do not see how the arguments of section 8 of
\cite{mns}, which seemed to furnish a shortcut
explanation of the SU$(N)$ results, could be adapted to reproduce 
the numbers highlighted in table 1.  

We next turn to numerical verification of these proposed bulk indices.

\begin{center}
Table 1: $D=4$ and $D=6$ bulk indices for the 
orthogonal, symplectic and exceptional
groups of rank~$\leq 3$\\
\vspace{0.5cm}
\begin{tabular}{||c|| c | c | c ||} \hline
Group  & rank  & ind$_0^{D=4,6}$ 
\\ \hline  \hline  
SO(3)  & 1 & $1/4$  \\
SO(4)  & 2 & $1/16$  \\
SO(5)  & 2 & ${\bf 9/64}$       \\
SO(6)  & 3 & $1/16$         \\
SO(7)  & 3 & ${\bf 25/256}$      \\ \hline \hline 
Sp(2)  & 1 & $1/4$  \\
Sp(4)  & 2 & ${\bf 9/64}$ \\
Sp(6)  & 3 & ${\bf 51/512}$      \\ \hline \hline 
G$_2$  & 2 & ${\bf 151/864}$    \\ \hline 
\end{tabular} 
\end{center}

\section{Monte Carlo evaluation of Yang-Mills Integrals}

As in previous works \cite{kns,ks1}, we  evaluate the Yang-Mills 
integrals using
Monte Carlo methods. Both the Pfaffian polynomial $\CP_{D,G}$ 
and the action $\CS = -\frac{1}{4 g^2} \Tr
[X_\mu,X_\nu] [X_\mu,X_\nu]$ in eq.(\ref{int}) are 
homogeneous functions of the  $X_{\mu}^A$
\begin{equation}
X_\mu^A \rightarrow \alpha X_\mu^A \; \; \; (\forall \mu\ ; A)\Rightarrow
\left\{ \begin{array}{ccl} 
 \CP(\{X_{\mu}^A\}) & \rightarrow & \alpha^{\frac{1}{2} \CN {\rm dim}(G)} \; 
\CP(\{X_{\mu}^A\})\\

                    &             &                   \\ 
 \CS(\{X_{\mu}^A\}) & \rightarrow & \alpha^4 \; \CS(\{X_{\mu}^A\})        
        \end{array}        
\right. 
.
\label{homogen} 
\end{equation}
We introduce polar coordinates $(X^1_1, \ldots, X^D_{{\rm
dim}(G)}) = (\Omega_d, R)$, with $d= D {\rm dim}(G)$ the
total dimension of the integral.  As an example, $ \CP(\Omega,1)$ and $
\CS(\Omega_d,1)$ denote the value of the Pfaffian polynomial and the action,
respectively, for a configuration $(X^1_1, \ldots, X^D_{{\rm
dim}(G)})$ on the surface of the $d-$dimensional unit hyper-sphere, 
with polar coordinates $(\Omega_d,R=1)$). 

Eq.(\ref{homogen}) allows us to to perform the $R$-integration 
analytically for
each value of  $\Omega_d$, and to express the
Yang-Mills integral as an expectation value over these angular variables:
\begin{equation}
\CZ_{D,G}^{\CN} = \frac { \int {\cal D}\Omega_d~z_{G}(\Omega_d) }
{ \int {\cal D}\Omega_d  },
\end{equation}
with
\begin{equation}
z_{G}(\Omega_d)=
2^{-d/2-1 }
\frac{ 
\Gamma\Big(\frac{d}{4}  + \frac{\CN }{8}{\rm dim}(G)\Big) }
{ \Gamma\Big(\frac{d}{2}\Big) } \times
\frac{ \CP(\Omega,1)}
{
\Big[ \CS(\Omega_d,1) \Big]^{\frac{d}{4} + \frac{\CN}{8}  {\rm dim}(G) }
}.
\label{havelz2}
\end{equation}

As discussed previously, the integrand $z_G(\Omega_d)$ is too singular
to be obtained by direct sampling of random points on the surface
of the unit hyper sphere.  Slightly modifying our procedure from
\cite{ks1}, we therefore write
\begin{equation}
\CZ_{D,G}^{\CN} =
\bigg[\frac{\int{\cal D}\Omega_d~z
\times z^{\alpha_1 - 1}}{\int{\cal D}\Omega_d~z} \bigg]^{-1}
\bigg[\frac{\int{\cal D}\Omega_d~z^{\alpha_1}
\times z^{\alpha_2 - \alpha_1}}{\int{\cal D}\Omega_d~z^{\alpha_1}} \bigg]^{-1}
\bigg[\frac{\int{\cal D}\Omega_d~z^{\alpha_2} 
\times z^{-\alpha_2 }}{\int{\cal D}\Omega_d~z^{\alpha_2}} \bigg]^{-1}.
\label{alphahop}
\end{equation} 
Each of the terms $[\;\;]$ in eq.(\ref{alphahop}) is computed in a separate
run. For example, the second quotient in eq.(\ref{alphahop}):
\begin{equation}
\frac{\int{\cal D}\Omega_d~z^{\alpha_1} 
\times z^{\alpha_2 - \alpha_1}}{\int{\cal D}\Omega_d~z^{\alpha_1}} = 
\bigg< z^{\alpha_2 - \alpha_1}  \bigg>_{\alpha_1},   
\label{expectation} 
\end{equation}
is simply the average of $z^{\alpha_2 - \alpha_1}$ for points
$\Omega_d$  on the unit hyper-sphere distributed according to the
probability distribution $\pi(\Omega_d) \sim z^{\alpha_1}(\Omega_d)$
(As it stands, eq.(\ref{expectation}) is immediately applicable
to $D=4$, where the integrand $z$ is positive semi-definite
\cite{kns,aabhn}. In the general case, we have to sample with
$|z^{\alpha_1}(\Omega_d)|$ (cf. \cite{kns})).

We sample angular variables $\Omega_d$ according to $z^{\alpha_1}$
with a Metropolis Markov-chain method, which we now explain:  At
each iteration of the procedure, two distinct indices $(A_1, \mu_1)$
and $(A_2, \mu_2)$, and an angle $0<\phi<2\pi$ are chosen randomly.
An unbiased trial move $\Omega_d \rightarrow \Omega'_d$ is then
constructed  by modifying solely the coordinates $X_{\mu_1}^{A_1}$
and $X_{\mu_2}^{A_2}$:
\begin{equation}
\left[ \begin{array}{c}
         X_{\mu_1}^{A_1} \\
                          \\ 
         X_{\mu_2}^{A_2}
        \end{array}
\right] 
\rightarrow 
\left[ \begin{array}{c}   
         X_{\mu_1}^{A_1} \\                              
                          \\ 
         X_{\mu_2}^{A_2}   
        \end{array}
\right]' 
= \sqrt{(X_{\mu_1}^{A_1})^2 + (X_{\mu_2}^{A_2})^2} 
\left[ \begin{array}{c}   
         \sin(\phi)   \\                              
                       \\ 
         \cos(\phi)
        \end{array}
\right],
\label{trial} 
\end{equation}
all other elements of $\{X_{\mu}^{A}\}$ remaining unchanged.  The
trial move eq.(\ref{trial}) preserves the norm $R$ of the vector
$\{X_{\mu}^{A}\}$, i. e. keeps the configuration on the surface of
the unit sphere.  Furthermore, it is unbiased (the probability to
propose  $\Omega_d \rightarrow \Omega'_d$ is the same as for the
reverse move).

Finally, the move  (for the example
in eq. (\ref{expectation})) is accepted according to the Metropolis
acceptance probability
\begin{equation}
P(\Omega     \rightarrow \Omega' ) = \min\bigg(1,\frac{ z^{\alpha_1}(\Omega')}
                                    {z^{\alpha_1}(\Omega)} \bigg).
\end{equation}

Empirically, we found the values  $\alpha_1=0.95, \alpha_2=0.6$ to
be appropriate.  Each of the averages in eq.(\ref{alphahop})  was
computed within between a few  hours and more than a thousand hours
of computer time (on a work station array), corresponding to a
maximum of $5 \times  10^9$ samples.  Results are presented in the
table below.

\begin{center}
Table 2: Direct evaluation of Yang-Mills integrals\\
\vspace{0.5cm}
\begin{tabular}
{||c||r@{ $\pm$ }l | c  ||} \hline
Group  & \multicolumn{2}{c|}{ Monte Carlo result}   & Exact  \\
$G$    & \multicolumn{2}{c|}{ $\CZ^{\CN=4}_{D=4,G}$}&        \\  \hline \hline
SO(3)  & 1.255   &0.003                             &     1.2533\ldots    \\
SO(4)  & 0.197   &0.004                             &     0.1963\ldots    \\
SO(5)  & 0.589   &0.004                             &     0.589\ldots    \\
SO(6)  & 0.0407  &0.0007                            &     0.04101\ldots   \\
SO(7)  & 0.0169  &0.0003                            &     0.01708\ldots   \\ 
\hline \hline
Sp(2)  & 1.253   &0.001                             &      1.2533\ldots   \\
Sp(4)  & 18.65   &0.2                               &     18.849\ldots   \\
Sp(6)  & 279.2   &9.7                               &     285.59\ldots \\ 
\hline \hline
G$_2$  & 6943    &120                               &     7011.4\ldots 
\\ \hline
\end{tabular}
\end{center}

Dividing the Monte Carlo results for $Z^{\CN=4}_{D=4,G}$ by the  corresponding
group volume factors
(cf. eqs (\ref{groupfacso}), (\ref{groupfacsp}), and (\ref{groupfacg})) 
we arrive 
at our numerical predictions for the bulk indices  ind$^{D=4}_0(G)$, which we
compare below to the proposed analytical values.  

\begin{center}
Table 3:  Monte Carlo results for the $D=4$ bulk index  \\
\vspace{0.5cm}
\begin{tabular}
{||c||r@{ $\pm$ }l | r@{  (}l  ||} \hline
Group  & \multicolumn{2}{c|}{ Monte Carlo}   & \multicolumn{2}{c||}
{ Exact}      \\
$G$    & \multicolumn{2}{c|}{ ind$^{D=4}_0(G)$}              & \multicolumn{2}
{c||}{ (Table 1)}  \\  \hline \hline SO(3)  & 0.2503  &0.0006
& 0.25     & 1/4)    \\ SO(4)  & 0.0627  &0.0013
& 0.0625   & 1/16)   \\ SO(5)  & 0.1406  &0.001
& 0.1406   & {\bf 9/64}) \\ SO(6)  & 0.0620  &0.001
& 0.0625   & 1/16)   \\ SO(7)  & 0.0966  &0.0017
& 0.0976   & {\bf 25/256}) \\ \hline \hline Sp(2)  & 0.2500  &0.0002
& 0.25     & 1/4)    \\ Sp(4)  & 0.139   &0.0015
& 0.1406   & {\bf 9/64}) \\ Sp(6)  & 0.0973  &0.003
& 0.0996   & {\bf 51/512}) \\ \hline \hline G$_2$  & 0.173   &0.003
& 0.1747   & {\bf 151/864})\\ \hline \end{tabular} \end{center}

Agreement between the Monte Carlo results and theory is excellent, 
both in cases where rigorous results are known (SO(3), SO(4), Sp(2))
and where the deformation technique was applied. Among the latter cases, 
we again indicate in bold type {\em new} values, which had not been 
obtained before. 

\section{Bosonic convergence for orthogonal, symplectic and
exceptional integrals} 

Our qualitative Monte Carlo method (cf. \cite{ks1},\cite{ks2},\cite{kps})
allows us to determine the convergence properties of the bosonic
Yang-Mills integrals $Z^{\CN=0}_{D,G}$.  We have found the
following:
\begin{equation}
\begin{array}{lll} 
\left. 
\begin{array}{l} 
\CZ_{D,{\rm SO}(N=3,4)}^{\CN = 0}\\ 
                        \\ 
\CZ_{D,{\rm Sp}(2)}^{\CN = 0}\\ 
\end{array}
\right\}                      &< \infty\;\;\;  {\rm for  }\;\;\; &  
D \ge 5    \\ 
                              &                                  &
             \\ 
\left. 
\begin{array}{l} 
\CZ_{D,{\rm SO}(N \ge  5)}^{\CN = 0}\\ 
                        \\ 
\CZ_{D,{\rm Sp}(4,6,\ldots)}^{\CN = 0}\\ 
                        \\ 
\CZ_{D,{\rm G}_2}^{\CN = 0}\\ 
\end{array}
\right\}                      &< \infty\;\;\;  {\rm for  }\;\;\; &
  D \ge 3    \\ 
\end{array}
\end{equation}
All other bosonic integrals diverge.  We thus obtain conditions which are
fully consistent with the group isomorphisms discussed in the appendix.  
Let us
note that we also performed the same qualitative computations for the 
susy integrals, 
as an important check of the convergence of the underlying Markov 
chains during the
simulation.

\section{Conclusions and outlook}

In this paper we provided further evidence that Yang-Mills integrals
encode surprisingly rich and subtle structures, which may prove to
have important bearings on gauge and string theory.

The chief result of the present paper was to demonstrate that
Yang-Mills integrals, as well as the methods to study them,  can
be naturally generalized from the previously studied special unitary
symmetries to other gauge symmetries. We numerically evaluated the
partition functions for all semi-simple gauge groups of rank $r
\leq 3$ and compared the results to conjectured exact values, which
were obtained by a generalization of certain contour integrals
derived from a supersymmetric deformation procedure. The connection
between the Yang-Mills integrals and the bulk indices is provided
by the group volumes, that we computed explicitly. We provided
details on these very subtle calculations.  Agreement between the
approaches is perfect within the tight error margins left by our
Monte Carlo technique.

It would be very interesting to gain a simpler understanding of
the rational numbers collected in table 1,  although it is already
evident that the bulk indices of the groups in question are more
complicated than those of the special unitary case. In particular
it would be nice to find  general formulas for arbitrary rank.

In the present paper we have focused on $D=4$ since this
is the case where our numerical approach is most accurate.
One should clearly study the dimensions $D=6$ and, especially,
$D=10$ as well. It is straightforward, if more involved, 
to work out the predictions of the deformation method for these cases, 
at least for low rank gauge groups.
In \cite{kasm} exact values for the {\it total} (bulk plus boundary)
Witten index ind$^D(G)$ were proposed. In $D=4,6$ one should have
ind$^{D=4,6}(G)=0$ while for $D=10$ it is argued to be a positive
integer which, for groups other than SU$(N)$, can be
larger than one.
It is important to check the arguments by computing the index from
the path integral. The results in the present paper indicate
that the bulk contributions ind$_0^D(G)$ are correctly reproduced
by the defomation method; however, we still lack
a reliable method for computing the boundary terms
ind$_1^D(G)$.

Since the deformation method of \cite{mns} successfully reproduces
the partition functions, it is natural to ask whether it can be
extended to calculate correlation functions of the ensembles
eq.(\ref{susyint}), such as the quantities studied numerically (so
far only for SU$(N)$) in \cite{ks2},\cite{kps},\cite{aabhn}.  This
would likely lead to new insights both
in string theory \cite{ikkt1} and gauge theory \cite{ek},\cite{aabhn}.

Numerically, it might be interesting to compare SU$(N)$, SO$(N)$
and Sp$(2 N)$ for large values of $N$, as in the standard large
$N\rightarrow \infty$ limit of `t~Hooft these groups are expected
to lead to identical results.
 
\acknowledgements
We thank H.~Nicolai, H.~Samtleben, G.~Schr\"oder and 
especially A.~Smilga for useful discussions.
This work
was supported in part by the EU under Contract FMRX-CT96-0012.

\appendix

\section{Details and conventions for ${\rm SO}(N)$}

The Lie algebra so$(N)$ has $\frac{1}{2} N (N-1)$ generators
which we choose to be the following standard anti-symmetric
matrices $T^{pq}$ (i.e.~the index $A$ becomes a double index $pq$)
\[
(T^{pq})_{jk} =\delta^p_j \delta^q_k-\delta^p_k \delta^q_j
\label{gen}
\]
where $p<q$ $(p,q=1,2,\ldots,N)$.
The Lie algebra reads then
\[
[T^{pq},T^{rs}]=\delta^{qr} T^{ps}-\delta^{qs} T^{pr} -\delta^{pr} T^{qs}+
\delta^{ps} T^{qr}=\sum_{t<u} f^{pq,rs,tu} T^{tu}.
\label{algebra}
\]
In this basis the generators are normalized as
$\Tr T^{pq} T^{rs}=-2 \delta^{pq,rs}$ and the structure constants
are given through
$f^{pq,rs,tu}=-\frac{1}{2} \Tr \Big( T^{pq} [ T^{rs},T^{tu}] \Big)$.
The gauge potentials are then $X_\mu=\sum_{p<q} X_\mu^{pq} T^{pq}$
and the SO$(N)$ Yang-Mills integrals in these conventions read
\begin{equation}
\CZ_{D,{\rm SO}(N)}^{\CN} = 
\int \prod_{p<q}^{\frac{1}{2}N (N-1)} \prod_{\mu=1}^{D}  
\frac{d X_{\mu}^{pq}}{\sqrt{2 \pi}}
\exp \bigg[  \frac{1}{8} \Tr
[X_\mu,X_\nu] [X_\mu,X_\nu] \bigg]~
\CP_{D,N}(X),
\label{sosusint}
\end{equation}
where $\CP_{D,N}$ is the Pfaffian as defined in eq.(\ref{pfaff}). 
These conventions are such that, in
view of the isomorphism ${\rm so}(3)={\rm su}(2)$ we have for all $\CN$ 
(i.e.~$\CN=0,4,8,16$)
\begin{equation}
\CZ_{D,{\rm SO}(3)}^{\CN}=\CZ_{D,{\rm SU}(2)}^{\CN}=\CZ_{D,{\rm Sp}(2)}^{\CN},
\label{a1dynkin}
\end{equation}
where the sympletic case is discussed in appendix B.
One checks that in these normalizations the isomorphism
${\rm so(4)}={\rm so}(3) \oplus {\rm so}(3)$ results in
\begin{equation}
\CZ_{D,{\rm SO}(4)}^{\CN}=
2^{-\frac{3}{2} (D-\frac{1}{2}\CN )} \Big( \CZ_{D,{\rm SO}(3)}^{\CN} \Big)^2,
\label{d2dynkin}
\end{equation}
while the isomorphism ${\rm so}(6)={\rm su}(4)$ leads to
\begin{equation}
\CZ_{D,{\rm SO}(6)}^{\CN}=
2^{-\frac{15}{4} (D-\frac{1}{2}\CN )}  \CZ_{D,{\rm SU}(4)}^{\CN},
\label{d3dynkin}
\end{equation}
where the SU$(4)$ integral is defined as in \cite{kns}. 
Finally, the isomorphism ${\rm so}(5)={\rm sp}(4)$ translates into
\begin{equation}
\CZ_{D,{\rm SO}(5)}^{\CN}=
2^{-\frac{5}{2} (D-\frac{1}{2}\CN )}  \CZ_{D,{\rm Sp}(4)}^{\CN},
\label{b2dynkin}
\end{equation}
where the Sp$(4)$ integral is defined in appendix B.
One verifies that these isomorphisms are in perfect agreement with
the results of the present paper as well as with \cite{ks1}.

We next provide the details necessary for verifying the
group volume factor $\CF_{{\rm SO}(N)}$ of eq.(\ref{groupfacso}).
The natural Cartan subalgebra is spanned by the generators
$T^{12},T^{34}, \ldots$. The corresponding maximal compact tori
are given for SO$(2N+1)$ by the $(2N+1) \times (2N+1)$ matrix 
\begin{equation}
T= \left(
\begin{array}{ccccc}
{\rm rot} \theta_1 
           &      &   &  &  \\
  & {\rm  rot} \theta_2    &   &  & \\
  &   &  \ddots  &    &\\
  &   &   &    {\rm  rot} \theta_N 
                        &         \\
  &   &   &       & {\large 1}\\
\end{array}
\right)
\end{equation}
while for SO$(2N)$ one has the $2N \times 2N$ matrix 
\begin{equation}
T=\left(
\begin{array}{cccc}
{\rm rot} \theta_1 
           &      &   &    \\
  & {\rm  rot} \theta_2    &   &   \\
  &   &  \ddots  &    \\
  &   &   &    {\rm  rot} \theta_N \\
\end{array}
\right)
\end{equation}
Here ${\rm rot} \theta_i$ are the $2 \times 2$ rotation matrices
\begin{equation}
{\rm  rot} \theta_i = 
\left(
\begin{array}{cc}
\cos \theta_i   & \sin \theta_i   \\
-\sin \theta_i  & \cos \theta_i   \\
\end{array}
\right)
\end{equation}
and matrix elements with no entries are zero. 
The corresponding reduced, normalized
Haar measure on SO$(2N+1)$ (i.e.~eq.(\ref{weyl})) reads
\begin{equation}
\CD T=\frac{2^{2 N^2}}{2^N N!} 
~\prod_{i=1}^N \frac{d \theta_i}{2 \pi}
\prod_{i<j}^N \sin^2 \Big( \frac{\theta_i-\theta_j}{2} \Big)
 \sin^2 \Big( \frac{\theta_i+\theta_j}{2} \Big)~
\prod_{i=1}^N \sin^2 \Big( \frac{\theta_i}{2} \Big)
\label{haarsoodd}
\end{equation}
while for SO$(2N)$ one has
\begin{equation}
\CD T=\frac{2^{2 N (N-1)}}{2^{N-1} N!} 
~\prod_{i=1}^N \frac{d \theta_i}{2 \pi}
\prod_{i<j}^N \sin^2 \Big( \frac{\theta_i-\theta_j}{2} \Big)
 \sin^2 \Big( \frac{\theta_i+\theta_j}{2} \Big).
\label{haarsoeven}
\end{equation}
Eqs.(\ref{haarsoodd}),(\ref{haarsoeven}) may also be used to work out
the detailed form of the contour integrals 
eqs.(\ref{soodd}),(\ref{soeven})
of section 4: One simply
expands the Haar measure around $\theta_i \sim 0$. 

Finally we recall the center groups of SO$(N)$: One has
$Z_{{\rm SO}(2 N+1)}=\{ \id \}$ and
$Z_{{\rm SO}(2 N)}=\{ \id, -\id \}$ and therefore
$|Z_{{\rm SO}(2 N+1)}|=1$ and $|Z_{{\rm SO}(2 N)}|=2$.

\section{Details and conventions for ${\rm Sp}(2N)$}

The Lie algebra sp$(2N)$ has $2 N^2+N $ generators which we choose as
follows. Define the $N \times N$ matrices $E^{pq}$ 
($p,q,j,k=1,2,\ldots ,N $)
\[
(E^{pq})_{jk}=\delta^p_j \delta^q_k
\] 
Then we define the $N \times N$ matrix
generators $T^{a,pq}$ for $p<q$ by
\[
T^{0,pq} = 
\frac{1}{2 \sqrt{2}}\left(
\begin{array}{cc}
E^{pq}-E^{qp}  & 0  \\
0 & E^{pq}-E^{qp}  \\
\end{array}
\right)
\qquad
T^{1,pq} = 
\frac{1}{2 \sqrt{2}}\left(
\begin{array}{cc}
0  & i(E^{pq}+E^{qp})  \\
i(E^{pq}+E^{qp}) & 0  \\
\end{array}
\right)
\]
\[
T^{2,pq} = 
\frac{1}{2 \sqrt{2}}\left(
\begin{array}{cc}
0  & (E^{pq}+E^{qp})  \\
-(E^{pq}+E^{qp}) & 0  \\
\end{array}
\right)
\quad
T^{3,pq} = 
\frac{1}{2 \sqrt{2}}\left(
\begin{array}{cc}
i(E^{pq}+E^{qp})  & 0  \\
0 & -i(E^{pq}+E^{qp})  \\
\end{array}
\right)
\]
and the remaining generators are ($p=1,\ldots, N$)
\[
T^{1,pp} = 
\frac{1}{2}\left(
\begin{array}{cc}
0  & i E^{pp}  \\
i E^{pp} & 0  \\
\end{array}
\right)
\]
\[
T^{2,pp} = 
\frac{1}{2}\left(
\begin{array}{cc}
0  & E^{pp}  \\
-E^{pp} & 0  \\
\end{array}
\right)
\quad
T^{3,pp} = 
\frac{1}{2}\left(
\begin{array}{cc}
i E^{pp}  & 0  \\
0 & -i E^{pp}  \\
\end{array}
\right)
\]
In this basis the generators are normalized as 
$\Tr T^A T^B=-\half \delta^{AB}$
and the structure constants are given through 
$f^{ABC}=-2 \Tr T^A [T^B,T^C]$ where $A$ is the multi-index $(a,pq)$.
The Cartan subalgebra is spanned by the generators
$T^{3,pp}$ with $p=1, \ldots, N$. The corresponding maximal compact torus
is given by the matrix 
\begin{equation}
T= \left(
\begin{array}{cccccc}
e^{i \theta_1} &      &   &  & & \\
 & \ddots & & & &
\\&    & e^{i \theta_N}   &  & & \\
  &   &    & e^{-i \theta_1}   & & \\
  &   &   &    & \ddots &        \\
  &   &   &       &  & e^{-i \theta_N} \\
\end{array}
\right)
\end{equation}
The corresponding normalized Haar measure reads
\begin{equation}
\CD T=\frac{2^{2 N^2}}{2^N N!} 
~\prod_{i=1}^N \frac{d \theta_i}{2 \pi}
\prod_{i<j}^N \sin^2 \Big( \frac{\theta_i-\theta_j}{2} \Big)
 \sin^2 \Big( \frac{\theta_i+\theta_j}{2} \Big)~
\prod_{i=1}^N \sin^2 \theta_i
\label{haarsp}
\end{equation}

The center of Sp$(2N)$ is the group $Z_{{\rm Sp}(2N)}=\{ \id, -\id \}$
and thus $|Z_{{\rm Sp}(2 N)}|=2$.

\section{Details and conventions for ${\rm G}_2$}

The Lie algebra $G_2$ has 14 generators. For the fundamental
representation we choose them as the following explicit
$7 \times 7$ matrices (see e.g.~\cite{pesando}). Define
\[
(X^{p,q})_{jk} =\delta^p_j \delta^q_k-\delta^p_k \delta^q_j
\label{gen}
\]
\[
(Y^{p,q})_{jk} =i(\delta^p_j \delta^q_k+\delta^p_k \delta^q_j)
\label{gen}
\]
Then
\begin{eqnarray}
T^1&=&(X^{1,2} + X^{3,4}) \frac{1}{2 \sqrt{6}} +(X^{5,6} + X^{6,7} ) 
\frac{1}{2 \sqrt{3}} \nonumber \\ 
T^2&=&(Y^{1,2} + Y^{3,4}) \frac{1}{2 \sqrt{6}} +(Y^{5,6} + Y^{6,7} ) 
\frac{1}{2 \sqrt{3}}\nonumber \\
T^3&=&(X^{1,7} - X^{4,5}) \frac{1}{2 \sqrt{2}} \nonumber \\
T^4&=&(Y^{1,7} + Y^{4,5}) \frac{1}{2 \sqrt{2}} \nonumber \\
T^5&=&(X^{1,6} + X^{4,6}) \frac{1}{2 \sqrt{3}} +(-X^{2,7} - X^{3,5} ) 
\frac{1}{2 \sqrt{6}}\nonumber \\
T^6&=&(Y^{1,6} - Y^{4,6}) \frac{1}{2 \sqrt{3}} +(-Y^{2,7} + Y^{3,5} ) 
\frac{1}{2 \sqrt{6}}\nonumber \\
T^7&=&(X^{1,3} + X^{2,4}) \frac{1}{2 \sqrt{2}} \nonumber \\
T^8&=&(Y^{1,3} + Y^{2,4}) \frac{1}{2 \sqrt{2}} \nonumber \\
T^9&=&(-X^{1,5} + X^{4,7}) \frac{1}{2 \sqrt{6}} +(X^{2,6} - X^{3,6} ) 
\frac{1}{2 \sqrt{3}}\nonumber \\
T^{10}&=&(-Y^{1,5} - Y^{4,7}) \frac{1}{2 \sqrt{6}} +(Y^{2,6} + Y^{3,6} ) 
\frac{1}{2 \sqrt{3}}\nonumber \\
T^{11}&=&(-X^{2,5} - X^{3,7}) \frac{1}{2 \sqrt{2}} \nonumber \\
T^{12}&=&(-Y^{2,5} + Y^{3,7}) \frac{1}{2 \sqrt{2}} 
\end{eqnarray}
Finally, the matrices $T^{13}$ and $T^{14}$ are diagonal matrices with 
elements:
\begin{equation}
T^{13} = {\rm diag}( 
\frac{i}{2 \sqrt{6}}; 
\frac{-i}{2 \sqrt{6}}; 
\frac{i}{2 \sqrt{6}}; 
\frac{-i}{2 \sqrt{6}}; 
\frac{i}{ \sqrt{6}}; 
0; 
\frac{-i}{ \sqrt{6}} )
\end{equation}
\begin{equation}
T^{14} = {\rm diag}( 
\frac{i}{2 \sqrt{2}}; 
\frac{i}{2 \sqrt{2}}; 
\frac{-i}{2 \sqrt{2}}; 
\frac{-i}{2 \sqrt{2}}; 
0; 
0; 
0) 
\end{equation}

In this basis the generators are normalized as 
$\Tr T^A T^B=-\half \delta^{AB}$ and the 
structure constants are given through 
$f^{ABC}=-2 \Tr T^A [T^B,T^C]$. 
The Cartan subalgebra is spanned by the generators
$T^{13}$ and $T^{14}$. The corresponding maximal compact torus
is given by the matrix 
\begin{equation}
T=\left(
\begin{array}{ccccccc}
e^{ i \theta_1}   &  &  &  &  &  &  \\
 &e^{-i \theta_2}  &  &  &  &  &  \\
 &  & e^{ i \theta_2} &  &  &  &  \\
 &  &  & e^{ - i \theta_1} &  &  &  \\
 &  &  &  & e^{ i( \theta_1 + \theta_2) } &  &  \\
 &  &  &  &  & 1 &  \\
 &  &  &  &  &  & e^{ - i (\theta_1 + \theta_2 )} \\
\end{array}
\right)
\end{equation}
The normalized Haar measure on G$_2$ with respect to this torus reads
\begin{equation}
\CD T=\frac{2^{12}}{12} 
~\frac{d \theta_1}{2 \pi} \frac{d \theta_2}{2 \pi}
\sin^2 \Big( \frac{\theta_1}{2} \Big)
\sin^2 \Big( \frac{\theta_2}{2} \Big)
\sin^2 \Big( \frac{\theta_1-\theta_2}{2} \Big)
\sin^2 \Big( \frac{\theta_1+\theta_2}{2} \Big)
\sin^2 \Big( \frac{2 \theta_1+\theta_2}{2} \Big)
\sin^2 \Big( \frac{\theta_1+2 \theta_2}{2} \Big)
\label{haarg}
\end{equation}

The center of G$_2$ is trivial: $Z_{{\rm G}_2}=\{ \id\}$
and thus $|Z_{{\rm G}_2} |=1$.


\begin{references}
%
\bibitem{smilga} A.V.~Smilga, {\it Witten Index Calculation in
Supersymmetric Gauge Theory}, Yad. Fiz. 42 (1985) 728,
Nucl. Phys. B266 (1986) 45; A.V.~Smilga, {\it Calculation of the
Witten Index in Extended Supersymmetric Yang-Mills Theory}, (in Russian)
Yad. Fiz. 43 (1986) 215.
%
\bibitem{sestern} P.~Yi, {\it Witten Index and Threshold Bound States
of D-Branes}, Nucl.~Phys.~B505 (1997) 307, {\tt hep-th/9704098}; 
S.~Sethi and M.~Stern, {\it D-Brane Bound State Redux}, 
Commun. Math. Phys. 194 (1998) 675, {\tt hep-th/9705046}.
%
\bibitem{hollo1} N.~Dorey, T.J.~Hollowood, V.V.~Khoze, M.P.~Mattis, 
and S.~Vandoren, {\it Multi-Instantons and Maldacena's Conjecture}, 
JHEP 9906 (1999) 023, {\tt hep-th/9810243};
{\it Multi-Instanton Calculus and the AdS/CFT Correspondence in} $\CN=4$
{\it Superconformal Field Theory}, 
Nucl.~Phys.~B552 (1999) 88, {\tt hep-th/9901128}.
%
\bibitem{hollo2} T.J.~Hollowood, V.V.~Khoze and M.P.~Mattis,
{\it Instantons in $\CN=4$ Sp$(N)$ and SO$(N)$ theories
and the AdS/CFT correspondence}, {\tt hep-th/9910118}. 
%
\bibitem{ikkt1} N.~Ishibashi, H.~Kawai, Y.~Kitazawa and A.~Tsuchiya,
{\it A Large N Reduced Model as Superstring}, Nucl. Phys. B498 (1997) 467,
{\tt hep-th/9612115}.
%
\bibitem{ek} T.~Eguchi and H.~Kawai,
{\it Reduction of dynamical degrees of freedom in the large N gauge theory},
Phys. Rev. Lett. 48 (1982) 1063.
%
\bibitem{aabhn} J.~Ambj{\o}rn, K.N.~Anagnostopoulos, W.~Bietenholz,
T.~Hotta and J.~Nishimura, {\it Large $N$ Dynamics of Dimensionally
Reduced 4D SU$(N)$ Super Yang-Mills Theory}, {\tt hep-th/0003208}.
%
\bibitem{kns} W.~Krauth, H.~Nicolai and M.~Staudacher,
{\it Monte Carlo Approach to M-Theory}, Phys. Lett. B431 (1998) 31, 
{\tt hep-th/9803117}.
%
\bibitem{ks1} W.~Krauth and M.~Staudacher,
{\it Finite Yang-Mills Integrals},
Phys. Lett. B435 (1998) 350,
{\tt hep-th/9804199}.
%
\bibitem{ks2} W.~Krauth and M.~Staudacher,
{\it Eigenvalue Distributions in Yang-Mills Integrals},
Phys. Lett. B453 (1999) 253,
{\tt hep-th/9902113}.
%
\bibitem{kps} W.~Krauth, J.~Plefka and M.~Staudacher,
{\it Yang-Mills Integrals},
Class. Quant. Grav. 17 (2000) 1171,
{\tt hep-th/9911170}. 
%
\bibitem{nishi} T.~Hotta, J.~Nishimura and A.~Tsuchiya,
{\it Dynamical Aspects of Large $N$ Reduced Models},
Nucl. Phys. B545 (1999) 543,
{\tt  hep-th/9811220 }  
%
\bibitem{kasm} V.G.~Kac and A.V.~Smilga,
{\it Normalized Vacuum States in $\CN=4$ Supersymmetric Yang-Mills
Quantum Mechanics with any gauge group},
{\tt hep-th/9908096}.
%
\bibitem{mns} G.~Moore, N.~Nekrasov and S.~Shatashvili,
{\it D-particle bound states and generalized instantons}, 
Commun. Math. Phys. 209 (2000) 77, 
{\tt hep-th/9803265}.
%
\bibitem{mehta}M.L.~Mehta, {\it Random Matrices, 2nd edition},
Academic Press 1991.
%
\bibitem{pesando} I.~Pesando {\it Exact Results for the 
Supersymmetric G$_2$ Gauge Theories},
Mod. Phys. Lett. A10 (1995) 1871,
{\tt hep-th/9506139}. 
%


\end{references}
\end{document}